\documentclass[twocolumn,showpacs,aps,epsfig]{revtex4}

\usepackage{graphicx}
\usepackage{epstopdf}
\usepackage{latexsym}

\usepackage[center]{subfigure}

\begin{document}

 \newcommand{\bq}{\begin{equation}}
 \newcommand{\eq}{\end{equation}}
 \newcommand{\bqn}{\begin{eqnarray}}
 \newcommand{\eqn}{\end{eqnarray}}
 \newcommand{\nb}{\nonumber}
 \newcommand{\lb}{\label}
\newcommand{\PRL}{Phys. Rev. Lett.}
\newcommand{\PL}{Phys. Lett.}
\newcommand{\PR}{Phys. Rev.}
\newcommand{\CQG}{Class. Quantum Grav.}

\title{No-Go Theorem in   Spacetimes
with Two Commuting Spacelike Killing Vectors}
\email{Anzhong_Wang@baylor.edu} 
\author{ Anzhong Wang}
\affiliation{ CASPER, Physics Department, Baylor University,
Waco, TX76798-7316, USA}

\date{\today }

\begin{abstract}

Four-dimensional Riemannian spacetimes with two commuting spacelike Killing
vectors are studied in Einstein's theory of gravity, and found that no outer
apparent horizons exist, provided that the dominant energy condition holds.

\end{abstract}




\maketitle

\section{Introduction}

The final fate of a collapsing massive star, after it has exhausted
its nuclear fuel, has been one of the outstanding problems in
classical relativity. Despite of numerous efforts over the last three
decades or so,  our understanding is still mainly limited to several
conjectures, such as, the cosmic censorship conjecture
\cite{Penrose} and the hoop conjecture \cite{Thorne}. To the
former,  many counter-examples have been found \cite{Joshi},
although  it is still not clear whether those particular solutions
are  {\em generic} \cite{HIN02}. To the latter, no counter-example has
been found yet in four-dimensional spacetimes, but it has been
shown recently that this is no longer the case in high dimensions \cite{NM01}.

Due to its (mathematical) complexity, the
studies of gravitational collapse have been mainly
restricted to spacetimes with spherical symmetry \cite{Joshi}.
This is a very ideal case and there have been many attempts to study
the problem with less symmetry, for example,
in the  spacetimes with axial symmetry \cite{Chop03},
in which only one spacelike Killing vector exists. However,
analytical studies of these spacetimes seem still far
beyond our reach. Therefore, the next case would be spacetimes with two
spacelike Killing vectors, a subject that will be considered in this Letter.

\section{Spacetimes with two commuting spacelike Killing vectors}

Specifically,  we consider a four-dimensional
Riemannian spacetime $({\cal{M}}, g)$  with a signature $- 2$, and assume that
throughout the whole spacetime there exist two commuting spacelike Killing
vectors, $\xi_{(2)}$ and $\xi_{(3)}$,
\bq
\lb{2.1}
\left[\xi_{(2)}, \xi_{(3)}\right] = 0.
\eq
In order to have our results as much applicable as possible, in this Letter
we shall not impose any conditions on the nature of the orbits of these
Killing vectors, so they can be either open or closed.
In addition, we shall also not impose any conditions in asymptotical
region(s) of the spacetime, such as, asymptotical flatness. Therefore, the theorem
to be given below is valid for all the spacetimes with two commuting spacelike Killing
vectors, including the ones with rotation that have been rarely studied so far.

Then, it can be shown that there exist coordinates, $x^{\mu}\; (\mu = 0, 1, 2, 3)$,
in which we have $ \xi_{(2)} = \partial_{x^{2}}$, $\;\xi_{(3)} = \partial_{x^{3}}$,
and the metric $g$ is independent of $x^{2}$ and $x^{3}$.
Since the two-surface ${\cal{S}}$ spanned by $\xi_{(2)}$ and $\xi_{(3)}$ is spacelike,
there exist two null directions orthogonal to ${\cal{S}}$. Let $n_{\pm}$   denote
these directions and satisfy the condition $g(n_{+},n_{-})  = 1$.
We assume that ${\cal{M}}$ is orientable, and   $n_{\pm}$  are future-pointing.
Because the metric coefficients are independent of $x^{2}$ and $x^{3}$, it can
be shown that the corresponding one-forms of the null vectors $n_{\pm}$ are
proportional to gradients \cite{Sz72},
$n_{\pm}^{\flat} = {N_{\pm}}^{-1} \nabla\left(x_{\pm}\right)$,
where the symbols ``$\flat$" and $\nabla$ denote, respectively,
the covariant dual and absolute derivative with respect to $g$.
$N_{\pm}$ are arbitrary functions of $x^{0}$ and $x^{1}$ only,
subject to $g(n_{+},n_{-})  = 1$.
Choosing  $x_{\pm}$ as the coordinates  $x^{0}$ and $x^{1}$,   we find that
\bq
\lb{2.3}
n_{\pm}^{\flat} = {N_{\pm}}^{-1} dx_{\pm}.
\eq
To have  $n_{\pm}$ future-pointing, we must require $N_{\pm} > 0$.
Note that such defined coordinates $x_{\pm}$ are unique up to
$x_{\pm} = f_{\pm}\left(\tilde{x}_{\pm}\right)$.
However, this gauge freedom does not affect the discussions to be presented below.
Instead, one can use it to regularize the metric so that it is free of
coordinate singularities. In the following we assume that this is
the case.  For the details, we refer readers to \cite{Hay94,Wang03b}.

On the other hand, assuming that $m$ is a complex null vector tangent to ${\cal{S}}$
and satisfies the conditions
$g(m,m) = 0, \; g(m,\bar{m}) = -1$, we find that
$z_{(a)} = \left(n_{+},\; n_{-},\; m,\; \bar{m}\right)$ $\; (a = 1, 2,3,4)$
form a null tetrad \cite{NP62}, $g = n_{+}{\otimes}n_{-}  - m{\otimes}\bar{m}$,
where an overbar denotes the complex conjugate.
The components of these null vectors in the chosen coordinates  are given by
\bqn
\lb{2.4}
n_{+}^{\mu} &=& \left(0,\; N_{-}, \; X^{2}, \; X^{3}\right),\nb\\
n_{-}^{\mu} &=& \left(N_{+}, \; 0, \; Y^{2}, \; Y^{3}\right),\nb\\
m^{\mu} &=& \left(0,\;  0, \; Z^{2}, \; Z^{3}\right),
\eqn
where $X^{i},\; Y^{i}$ and $Z^{i}\; (i = 2, 3)$ are functions of $x_{\pm}$ only.
The components $X^{i}$ and $Y^{i}$ correspond to rotation \cite{Sz72}, while
the ones $Z^{i}$ to the two degrees of polarization of gravitational waves
  \cite{Wang91}.

Introducing the  one-forms $\hat{n}_{\pm}$ via the relations,
$\hat{n}_{\pm} \equiv N_{\pm} n_{\pm}^{\flat}$,
from Eq.(\ref{2.3}) we can see that $\hat{n}_{\pm}$ are closed,
$d\left(\hat{n}_{\pm}\right) = 0$.
Then, we define the expansions in the  null directions orthogonal to
${\cal{S}}$ by \cite{HE73,Note}
\bq
\lb{2.4a}
\theta_{\pm} \equiv  \nabla\cdot{\hat{n}_{\pm}}.
\eq

On the other hand, using the commutation relations  \cite{NP62},
we find that the spin coefficients have the following properties,
\bqn
\lb{2.5}
&& \kappa = \nu = 0,\;\;\; \rho = \bar{\rho},  \;\;\; \mu = \bar{\mu}, \nb\\
&& \alpha = \bar{\beta},\;\;\; \pi = \bar{\tau} = 2 \alpha,\nb\\
&& {\cal{R}}e\left(\epsilon\right)
= -  {\left(2N_{+}\right)}^{-1} D_{+}N_{+},\nb\\
&& {\cal{R}}e\left(\gamma\right)
  =  {\left(2N_{-}\right)}^{-1} D_{-}N_{-},
\eqn
where $D_{\pm} \equiv n_{\pm}\cdot{\nabla}$.
Since $\kappa = 0$, the null vector $n_{+}$  defines a null geodesic
congruence \cite{Chandra83}. Choosing $N_{+} = 1$, from Eq.(\ref{2.5})
we can see that ${\cal{R}}e\left(\epsilon\right) = 0$, and consequently
$n_{+}$ also defines an affine parameter,
say, $\lambda_{+}$, in terms of which we have
${\nabla \left(n_{+}\right)}/{\nabla \lambda_{+}} = 0$.
Then, the expansion $\theta_{+}$ defined by  Eq.(\ref{2.4a}) is given by
\bq
\lb{2.7a}
\theta_{+} \equiv    \nabla\cdot{\hat{n}_{+}}
= - 2 \left.\rho\right|_{N_{+} = 1}.
\eq
Replacing $\kappa, \epsilon, \rho$ by $-\nu, -\gamma,
-\mu$ in the above discussions, we can get the geometrical
properties of the null geodesic congruence defined by $n_{-}$.

{\em Definitions} \cite{Pen68,Hay94}:  The spatial two-surface ${\cal{S}}$
is said {\em trapped, marginally trapped, or untrapped}, according to whether
$\left. \theta_{+}\theta_{-}\right|_{{\cal{S}}} > 0$,
$\; \left. \theta_{+}\theta_{-}\right|_{{\cal{S}}} = 0$,
or $\left. \theta_{+}\theta_{-}\right|_{{\cal{S}}} < 0$.
Assuming that on the marginally trapped surfaces ${\cal{S}}$ we have
$\left.\theta_{+}\right|_{{\cal{S}}} = 0$, then an {\em apparent horizon} is
the closure $\tilde{\Sigma}$ of a three-surface $\Sigma$ foliated by the trapped
surfaces $\cal{S}$ on which $\left.\theta_{-}\right|_{\Sigma} \not= 0$.
It is said {\em outer, degenerate, or inner}, according to
whether $\left.{\cal{L}}_{-}\theta_{+}\right|_{\Sigma} < 0$,
$\left.{\cal{L}}_{-}\theta_{+}\right|_{\Sigma} = 0$, or
$\left.{\cal{L}}_{-}\theta_{+}\right|_{\Sigma} > 0$, where ${\cal{L}}_{-}$
denotes the Lie derivative along the normal direction
${n}_{-}$. In addition, if $\left. \theta_{-}\right|_{\Sigma} < 0$
then the apparent horizon is said {\em future}, and if
$\left. \theta_{-}\right|_{\Sigma} > 0$ it is said {\em past}.

{\em Black holes} are usually defined by the existence of {\em future outer
apparent horizons} \cite{HE73,Hay94,Ida00}. However, in a definition given by
Tipler \cite{Tip77} the degenerate case was also included, as first noted
by Hayward \cite{Hay94}.

In the following, we shall show that with a positive cosmological constant,
both outer and degenerate apparent horizons do not exist in the spacetimes
considered here. To this end, let us first notice that Eqs.(4.2q) and (4.2l)
in \cite{NP62} now read
\bqn
\lb{2.8a}
D_{-}\rho &=& \rho\left(\gamma + \bar{\gamma}\right) - \rho\mu - \sigma\lambda
               - \tau\bar{\tau} - \Psi_{2} - R/12,\\
\lb{2.8b}
0 &=& \rho\mu - \sigma\lambda -  \Psi_{2} + R/24 + \Phi_{11},
\eqn
where $R$ denotes the Ricci scalar, $\Psi_{2}$ and $\Phi_{11}$ are defined as
$\Phi_{11} \equiv - \left(n_{+}^{\mu}n_{-}^{\nu}
+ m^{\mu}\bar{m}^{\nu}\right)R_{\mu\nu}/4$, and
$\Psi_{2} \equiv  - C_{\alpha\beta\delta\sigma}
       n_{+}^{\alpha}m^{\beta}\bar{m}^{\delta}n_{-}^{\sigma}$,
where $C_{\alpha\beta\delta\sigma}$ is the Weyl tensor. From Eqs.(\ref{2.8a})
and (\ref{2.8b}) we find that
\bq
\lb{2.10}
D_{-}\rho = \rho\left(\gamma + \bar{\gamma} - 2\mu\right)
               - \tau\bar{\tau} - \left(R + 8\Phi_{11}\right)/8.
\eq
Choosing the particular gauge $N_{+} = 1$ and considering the fact that on the
apparent horizon we have $\left.\theta_{+}\right|_{\Sigma} = 0$, from
Eqs.(\ref{2.7a}) and (\ref{2.10})  we find that
\bq
\lb{2.11}
\left.{\cal{L}}_{-}\theta_{+}\right|_{\Sigma}
=   2\tau\bar{\tau}
+ \left(R + 8\Phi_{11}\right)/4,
\eq
on $\Sigma$. Clearly, the first term in the right-hand of the above equation 
is always non-negative. To consider the signs of the second term, following
Hawking and Ellis \cite{HE73}, we shall express the components of the 
energy-momentum tensor $T_{\mu\nu}$ at a given point $p$ with respect 
to an orthonormal basis $E_{(a)},\; (a = 1, 2, 3, 4)$, where
\bqn
\lb{2.11a}
E_{(4)} &\equiv& \frac{n_{+} + n_{-}}{\sqrt{2}},\;\;\;
E_{(3)} \equiv \frac{n_{+} - n_{-}}{\sqrt{2}},\nb\\
E_{(2)} &\equiv& \frac{m  + \bar{m}}{\sqrt{2}},\;\;\;
E_{(1)} \equiv \frac{m  - \bar{m}}{i \sqrt{2}}.
\eqn
Then, as shown in \cite{HE73}, it takes four different canonical forms, which
were referred to as Type I - IV, respectively. Types III and IV don't satisfy any
of the three energy conditions (weak, dominant, and strong), and usually are
not considered to represent realistic matter. For Types I and II,
using the Einstein field equations, $Ric - (R/2)g + \Lambda{g} = - {T}$, 
we find that
\bq
\lb{2.11b}
\left(R + 8\Phi_{11}\right) - 4\Lambda = 
\cases{2(\mu - p_{i}), & Type I,\cr
4k, & Type II,\cr}
\eq
where $i = 1, 2, 3$. Note that in writing the above expressions we had
considered the fact that the roles of the three spacelike vectors, 
$E_{(i)}$, can be exchanged. 

{\em The dominant energy condition}  requires that $\mu \ge 0,\; 
- \mu \le p_{i} \le \mu \; (i = 1, 2, 3)$ for Type I fluid, and
$\nu = +1,\; k \ge 0, \; 0 \le p_{j} \le k\; (j = 1, 2)$
for Type II fluid \cite{HE73}. Then, combining Eqs.(\ref{2.11}) 
and (\ref{2.11b}) we have the following:
 
{\bf Theorem}: {\em Let (${\cal{M}}, g$) be a four-dimensional Riemannian spacetime to the
Einstein field equations, $Ric - (R/2)g + \Lambda{g} = - {T}$ with
$\Lambda > 0$. Assume that throughout the spacetime there exist two commuting spacelike
Killing vectors, $[\xi_{(2)}, \xi_{(3)}] = 0$. Then,  (${\cal{M}}, g$) contains neither
outer nor degenerate apparent horizons, if the dominant energy condition holds}.

Note that when $\Lambda = 0$, the dominant energy condition only guarantees that
$R + 8\Phi_{11} \ge 0$, which together with Eq.(\ref{2.11}) implies that
$\left.{\cal{L}}_{-}\theta_{+}\right|_{\Sigma} \ge 0$. Therefore, in this case
only the existence of outer apparent horizons is excluded.

The significance of the above theorem  is two-fold. First, for a stationary spacetime,
a spacetime that has an additional timelike Killing vector (at least in certain
region(s) of the spacetime), say, $\xi_{(0)}$, with $g(\xi_{(0)}, \xi_{(0)}) > 0$,
then the above theorem tells us that {\em no black hole  exists,
unless $\Lambda \le 0$}.  This is consistent with the fact that so far all
the black holes with different topologies rather than that of $S^{2}$
\cite{Haw72} are with $\Lambda < 0$ \cite{TBHs}. This is also in the same spirit
of {\em topological censorship} \cite{FSW93}. It is interesting to note that so
far degenerate stationary black holes have not been found in the spacetimes
considered here, where ``degenerate" means that the future apparent horizon that
defines the black hole is degenerate.

Second, in the process of gravitational collapse of a source that satisfies the dominant
energy condition and has two commuting spacelike Killing vectors, the theorem tells us that
{\em black holes can never be formed, unless a negative cosmological constant is present,
$\Lambda < 0$}. In the case where $\Lambda = 0$, at most a ``degenerate" black hole can be
formed by the collapse. An example of such a dynamical ``degenerate" black hole
was found lately  in the study of critical collapse of a
cylindrically symmetric scalar field \cite{Wang03a}. Restricting ourselves to
these spacetimes, we can see that the above theorem supports the hoop conjecture
\cite{Thorne}: {\em Horizons form when and only when a mass $M$ gets  compacted into a
region whose circumference ${\cal{C}}$ in {\em every} direction is ${\cal{C}}
\le 4\pi GM/c^{2}$}. It should be noted that in \cite{BCM95} the ``asymptotical flatness"
condition was imposed in the spacetimes with cylindrical symmetry, and found that
no trapped surfaces can be formed in electro-vacuum case. This does not contradict
with the above theorem. In fact, when this condition is relaxed, degenerate
apparent horizons indeed exist in vacuum spacetimes \cite{Wang03b}. It should be
noted that the notion of asymptotical flatness in this kind of spacetimes is a very
delicate issue. For the details, we would like to refer readers to \cite{Ashtekar}.

To  study  the problem further,  let us consider the case where the two Killing vectors
all have closed orbits. Then, from the above theorem we can see that the collapse is more
likely to form naked singularities than ``degenerate" black holes for all the matter
fields that satisfy the dominant energy conditions with $\Lambda \ge 0$.  In fact,
compacting the axial coordinate  in the examples studied in \cite{PW00} for cylindrical
collapse, we can see that the resultant spacetimes  can be asymptotically flat,
but naked singularities may still be formed.

Spacetimes where not only the two-spaces ${\cal{S}}$ are compact but also all the
spatial hypersurfaces are compact have been intensively studied recently \cite{Ber02},
after the pioneering work of Gowdy \cite{Gowdy}. This kind of spacetimes is
usually divided into three different classes, according to the topologies of
these spatial hypersurfaces, ${\Sigma_{t}} \approx  T^{3} \equiv S^{1}\times
S^{1}\times S^{1}$, $\; S^{2}\times S^{1}$ or $S^{3}$,  where $S^{n}$ denotes
a unit $n$-sphere. In the case ${\Sigma_{t}} \approx
T^{3}$,   one can show that {\em the whole spacetime
$({\cal{M}}, g)$ developed from a Cauchy data on a compact hypersurface
$\left.{\Sigma_{t}}\right|_{t = t_{0}< 0}$ is trapped}, where $t$ is a
timelike coordinate and the spacetime is foliated by $t = Const$. In fact,
in this case one can show that the following holds in the entire spacetime
\bq
\lb{3.1}
\theta_{+}\theta_{-} = {{\cal{R}}^{-2}}e^{2f} {F'}_{+}{F'}_{-} > 0,
\eq
where $F_{\pm}[\equiv F_{\pm}(x_{\pm})]$ are arbitrary functions of their
indicated arguments, subject to  ${F'}_{+}{F'}_{-} > 0$, so that the coordinates
are future-pointing. A prime denotes the ordinary derivative,
$e^{f} \equiv g(\hat{n}_{+},\hat{n}_{-})$, and
\bq
\lb{3.2a}
{\cal{R}} \equiv \left|{\mbox{det}}\left(g(\xi_{(2)}, \xi_{(3)})\right)\right|^{1/2}
= F_{+} + F_{-}.
\eq
Thus, no apparent horizons exist in these spacetimes.
  In the cases ${\Sigma_{t}} \approx
S^{2} \times S^{1}$ or $ S^{3}$,  we have \cite{Ber02}
\bq
\lb{3.2b}
{\cal{R}}  = \sin{(t)}\sin{(\theta)},
\eq
with $t \equiv F_{+} + F_{-},\; \theta \equiv F_{+} - F_{-}$,
and $0 \le t, \; \theta \le \pi$. Again, to have the coordinates future-pointing,
we must have ${F'}_{+}{F'}_{-} > 0$. Then, one can show that now we have
\bq
\lb{3.3}
\theta_{\pm} =  {\cal{R}}^{-1}e^{f} {F'}_{\pm}\sin(\theta \pm t).
\eq
Thus, in this case the spacetime has several trapped   regions [cf. Fig.1].
However, it can be shown that {\em the apparent horizons that
separate the trapped  regions from the untrapped ones are  all degenerate}.
As a matter of fact, from Eq.(\ref{3.3}) we find
\bq
\lb{3.4}
{\cal{L}}_{\mp}\theta_{\pm} = {\cal{R}}^{-1}e^{f}\left({\cal{R}} f_{,\mp}
- {\cal{R}}_{,\mp}\right)
\theta_{\pm},
\eq
where $f_{,\mp} \equiv \partial f/\partial x_{\mp}$.
Thus,  at least one of the conditions ${\cal{L}}_{\mp}\theta_{\pm} = 0$
holds on the apparent horizons denoted by the lines $ad$ and $bc$ in Fig. 1,
on which we have $\theta_{-}\theta_{+} = 0$.
Therefore, in this case all these horizons are degenerate, which is consistent
with the above theorem.

\begin{figure}
\centering
\includegraphics[width=8cm]{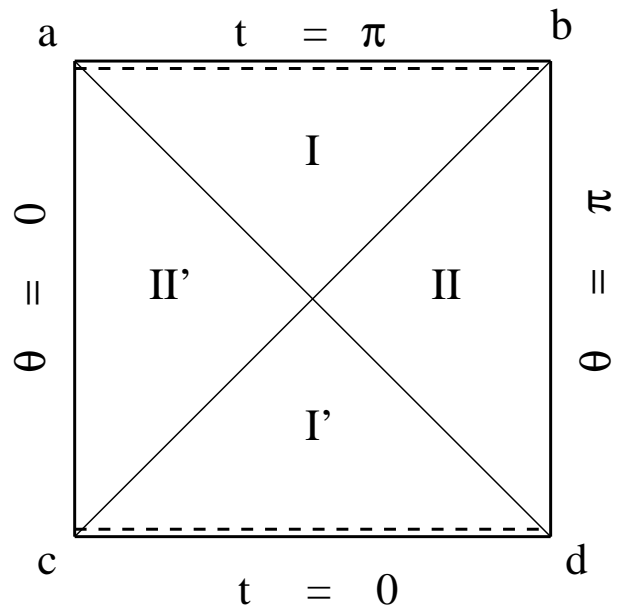}
\caption{The spacetime for the polarized Gowdy solutions in the ($t, \theta$)-plane
with the spatial topology ${\Sigma_{t}} \approx   S^{2} \times S^{1}$ or $ S^{3}$.
In the regions $I$ and $I'$ the two-surfaces of constant $t$ and $\theta$ are
untrapped ($\theta_{+}\theta_{-} < 0$), while in the regions $II$ and $II'$
they are trapped ($\theta_{+}\theta_{-} > 0$). The lines $ad$ and $bc$
are degenerate apparent horizons ($\theta_{+}\theta_{-} = 0,\;
{\cal{L}}_{-}\theta_{+} {\cal{L}}_{+}\theta_{-} = 0$), and the spacetime is
singular on the horizontal lines $t = 0, \; \pi$.}
\label{fig1}
\end{figure}

\section{ Conclusions} 

In this Letter, we have studied four-dimensional spacetimes
with two commuting spacelike Killing vectors and the cosmological constant
$\Lambda$. After defining trapped surfaces and apparent horizons, following 
Penrose \cite{Pen68} and Hayward \cite{Hay94}, we have been able to show that 
outer apparent horizons do not exist in such spacetimes with $\Lambda > 0$.
Degenerate apparent horizons can be formed only in the cases where 
$\Lambda \le 0$. These are consistent with all the results obtained so 
far in the studies of both stationary black holes 
\cite{TBHs,FSW93,SM97} and gravitational collapse \cite{Wang03a,PW00,Ber02}. 
In particular, it supports the hoop conjecture \cite{Thorne}.

\section*{Acknowledgments}

The author  would like to thank Nigel Goldenfeld  for his kind invitation,
and the Physics Department, UIUC, for hospitality. He also thanks Rong-Gen Cai
for valuable discussions and suggestions.


\begin{thebibliography}{100}
\bibitem{Note} It should be noted that the definitions adopted
 here are different from that used by Hayward \cite{Hay94}. In
 particular,  it can be shown that $\theta_{\pm} \propto
 \theta_{\mp}^{Hayward}$.
\end{thebibliography}
\end{document}